\documentclass[aps,prl,twocolumn,superscriptaddress,showpacs,showkeys]{revtex4}

\usepackage{graphicx}
\usepackage{amsmath}

\newcommand{\ket}[1]{\mbox{$\,\mid \! #1 \, \rangle$}}
\newcommand{\bra}[1]{\mbox{$\langle \, #1 \! \mid \,$}}

\newcommand{\dg}[1]{#1 \,^{\circ}}

\newcommand{\dicke}[2]{\ket{D_{#1}^{(#2)}}\,}

\newcommand{\Dicke}[2]{D_{#1}^{(#2)}\,}

\newcommand{\cref}[1]{chapter~\ref{#1}}

\newcommand{\fref}[1]{Fig.~\ref{#1}}

\begin{document}


\title{Experimental entanglement of a six-photon symmetric Dicke state}


\author{Witlef Wieczorek}
\email[]{witlef.wieczorek@mpq.mpg.de}
\author{Roland Krischek}
\author{Nikolai Kiesel}
\author{Patrick Michelberger}
\affiliation{Max-Planck-Institut f{\"u}r Quantenoptik, Hans-Kopfermann-Strasse 1, D-85748 Garching, Germany}
\affiliation{Department f\"ur Physik, Ludwig-Maximilians-Universit{\"a}t, D-80797 M{\"u}nchen, Germany}
\author{G\'eza T\'oth}
\affiliation{IKERBASQUE and Department of Theoretical Physics, The University of the Basque Country, P.O.~Box 644, E-48080 Bilbao, Spain}
\affiliation{Research Institute for Solid State Physics and Optics, Hungarian Academy of Sciences, P.O.~Box 49, H-1525 Budapest, Hungary}
\author{Harald Weinfurter}
\affiliation{Max-Planck-Institut f{\"u}r Quantenoptik, Hans-Kopfermann-Strasse 1, D-85748 Garching, Germany}
\affiliation{Department f\"ur Physik, Ludwig-Maximilians-Universit{\"a}t, D-80797 M{\"u}nchen, Germany}



\begin{abstract}

We report on the experimental observation and characterization of a six-photon entangled Dicke state. We obtain a fidelity as high as $0.654\pm0.024$ and prove genuine six-photon entanglement by, amongst others, a two-setting witness yielding $-0.422\pm0.148$. This state has remarkable properties, e.g., it allows to obtain inequivalent entangled states of a lower qubit number via projective measurements and it possesses a high entanglement persistency against qubit loss. We characterize the properties of the six-photon Dicke state experimentally by detecting and analyzing the entanglement of a variety of multi-partite entangled states. 

\end{abstract}

\pacs{03.67.Bg, 03.65.Ud, 03.67.Mn, 42.50.Ex, 42.65.Lm}









\maketitle




Multi-partite entangled states have been intensively studied during recent years. Still, the experimental realization of entangled states of more than four particles imposes a considerable challenge and only few experiments have yet demonstrated such states \cite{Lei05,Haf05}. So far, many experiments focused on the observation of graph states \cite{Hei04} like the Greenberger-Horne-Zeilinger (GHZ) states or the cluster states \cite{Lei05}, which are, e.g., useful for one-way quantum computation \cite{Rau01}. Dicke states form another important group of states, which have first been investigated with respect to light emission from a cloud of atoms \cite{Dic54} and have now come into the focus of both experimental realizations \cite{Haf05,Kie93,Eib04,Kie07} and theoretical studies \cite{Se03b,Tot07,Tas08,Wie09}. W-states \cite{Dur00}, a subgroup of the Dicke states, first received attention triggered by the seminal work on three-qubit classification based on stochastic local operations and classical communication (SLOCC) by D\"ur {\emph{et al.}}~\cite{Dur00}. Recently it turned out that also other symmetric Dicke states offer important features. Particularly, by applying projective measurements on a few of their qubits, states of different SLOCC entanglement classes are obtained \cite{Kie07,Wie09}. These Dicke states can act as a rich resource of multi-partite entanglement as required for quantum information applications.

In our work we experimentally implement and analyze a symmetric six-qubit entangled Dicke state. The entanglement of the Dicke state results from symmetrization and cannot be achieved in a simple way by pairwise interaction, in contrast to, e.g., GHZ states. In order to efficiently characterize the experimentally observed state we developed optimized methods to determine the fidelity, detect entanglement and characterize further properties. In particular, we analyze representatives from the variety of multi-partite entangled states obtained after projection or loss of qubits.

Generally, Dicke states are simultaneous eigenstates of the total angular momentum, $J^2_N=J_{N,x}^2+J_{N,y}^2+J_{N,z}^2$, and the angular momentum component in the $z$ direction, $J_{N,z}$, where $J_{N,i}=1/2\sum_k{\sigma_i}^k$ with, e.g., ${\sigma_i}^3=\openone\otimes\openone\otimes\sigma_i\otimes\openone\otimes\openone\otimes\openone$ for $N=6$ qubits, $i\in\{x,y,z\}$ and $\sigma_i$ the Pauli spin matrices. A subgroup of the Dicke states are symmetric under permutation of particles and given by
\begin{equation} 
\dicke{N}{l}=\left(\begin{array}{c}N\\l\end{array}\right)^{-1/2}\sum_i\mathcal{P}_i(\ket{H^{\otimes (N-l)}V^{\otimes l}}),
\end{equation}
where $\sum_i\mathcal{P}_i(...)$ means the sum over all distinct symmetric permutations and $l$ is the number of excitations in the usual notation of polarization encoded photonic qubits. In our experiment we focus on the symmetric six-qubit Dicke state with three excitations,
\begin{equation} 
\dicke{6}{3}=1/\sqrt{20}\sum_i\mathcal{P}_i(\ket{HHHVVV}).
\end{equation}
To realize the necessary 20 permutations, three horizontally and three vertically polarized photons in a single spatial mode are distributed by polarization-independent beam splitters into six modes, where $\dicke{6}{3}$ is observed under the condition of detecting a single photon in each of these modes. This scheme can be seen as a continuation of experiments on $\Dicke{2}{1}$ \cite{Kie93} and $\Dicke{4}{2}$ \cite{Kie07} and obviously can be extended to higher even photon numbers.

\begin{figure}
\includegraphics{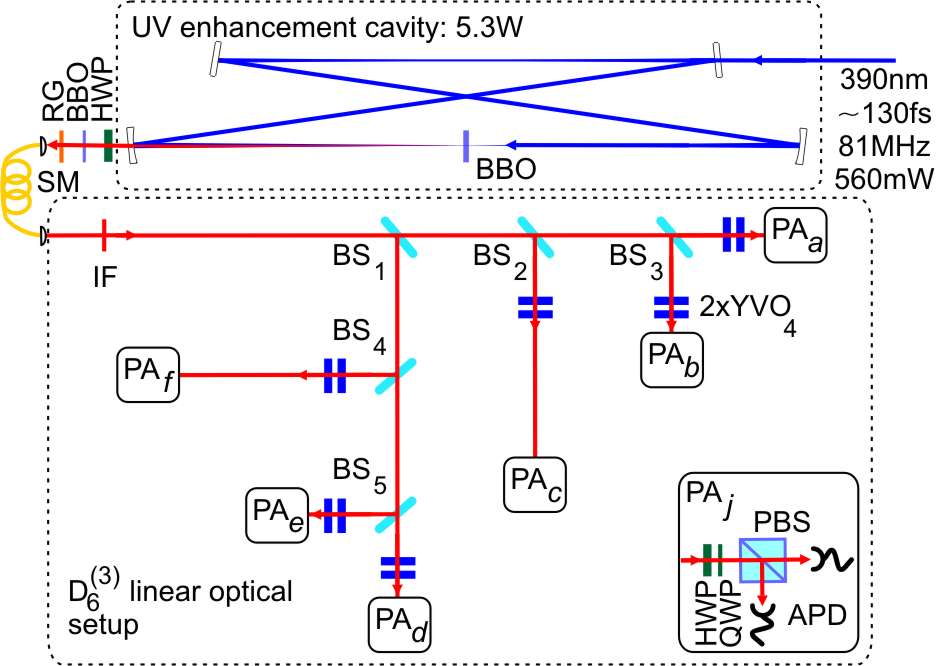}
\caption{\label{fig:setup}(color online) Schematic experimental setup for the observation of the Dicke state \dicke{6}{3}. SPDC photons generated in the 1\,mm thick $\beta$-barium borate (BBO) crystal inside the UV enhancement cavity pass a half-wave plate (HWP) and a 0.5\,mm thick BBO crystal to compensate beam walk-off effects. Their spatial mode is defined by coupling into a single-mode (SM) fiber. Spectral selection is achieved by a band-pass filter (RG) and a 3\,nm interference filter (IF) at 780\,nm. Birefringence of beam splitters BS$_1$-BS$_5$ (BS$_1$-BS$_4$ have a splitting ratio of $0.58:0.42$ and BS$_5$ of $0.52:0.48$) is compensated for by pairs of birefringent Yttrium-vanadate (YVO$_4$) crystals in the six output modes $a,b,c,d,e,f$. Polarization analysis (PA$_j$) in each mode is performed via a HWP and a quarter-wave plate (QWP) in front of a polarizing beam splitter (PBS). The photons are detected by single-photon avalanche photo-diodes (APD). The detection signals of the twelve detectors are fed into a FPGA controlled coincidence logic allowing histograming of the $2^{12}$ possible detection events between the twelve detectors.}
\end{figure}

The experimental observation of $\dicke{6}{3}$ (\fref{fig:setup}) is achieved by utilizing a novel source of collinear type II spontaneous parametric down conversion (SPDC) based on a femto-second UV-enhancement resonator \cite{Kri09tbd}. The resonator allows to pump the SPDC crystal with femto-second pulses with an average UV power of $5.3$\,W at a repetition rate of 81\,MHz \cite{Kri09tbd}. The SPDC photons are coupled out of the cavity by a dichroic mirror transparent at 780\,nm, are spatially filtered by a single-mode fiber and are subsequently distributed in free-space by polarization-independent beam splitters. Asymmetry in the splitting ratios of the beam splitters reduces the probability of registering $\dicke{6}{3}$ ($0.0126$ compared to the optimal value of $5/324\approx0.0154$, yielding a six-photon count rate of 3.7 events per minute), but does not influence the state quality. For all data the errors are deduced from Poissonian counting statistics and errors of independently determined relative detector efficiencies.

\begin{figure*}
\includegraphics{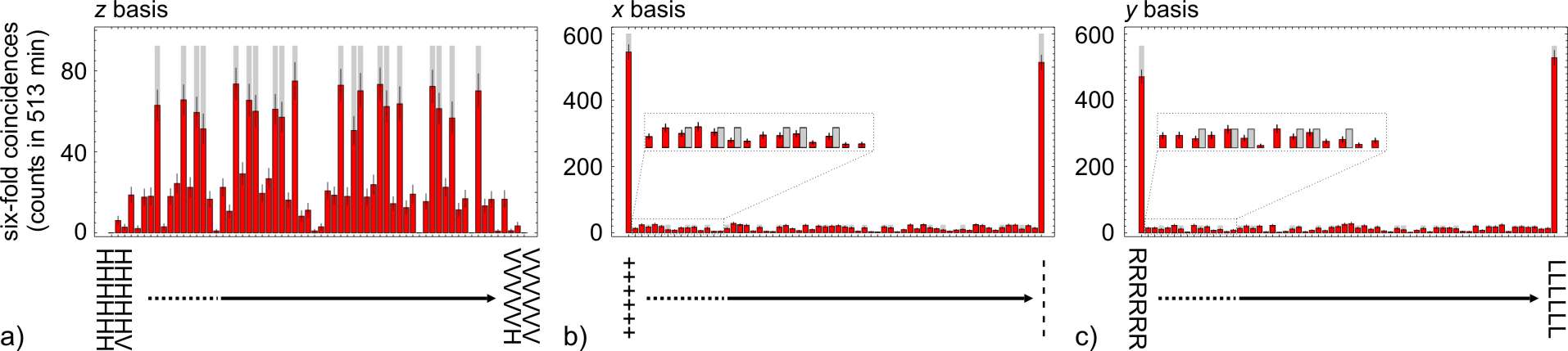}
\caption{\label{fig:counts}(color online) Experimentally measured coincidences for the bases (a) $z$, (b) $x$ and (c) $y$ with eigenvectors $\ket{H/V}$, $\ket{\pm}$ and $\ket{L/R}$, respectively. Theoretical predictions are shown as gray bars normalized to the total number of coincidences. The insets in (b) and (c) are magnified views of a part of all coincidences, where for clarity expected counts are shown next to experimental ones.}
\end{figure*}

The first characteristic feature of the state $\dicke{6}{3}$ is its structure in the $z$, $x$ and $y$ bases (\fref{fig:counts}), i.e., when analyzing the photons in the six outputs all either along $\ket{H/V}$, $\ket{\pm}=1/\sqrt{2}(\ket{H}\pm\ket{V})$ (linear polarization under $\dg{45}$) and $\ket{L/R}=1/\sqrt{2}(\ket{H}\pm i\ket{V})$ (left/right circular polarization), which, in our notation, are the eigenvectors of $\sigma_z$, $\sigma_x$ and $\sigma_y$, respectively. For the $z$ basis [\fref{fig:counts}(a)] we experimentally find the pronounced 20 terms that are expected for $\dicke{6}{3}$. However, we also detect coincidences for $HHVVVV$, $HHHHVV$ and permutations thereof. These originate from higher orders of the SPDC process, in particular, from the fourth order emission, where, due to the finite overall detection efficiency, two of these photons can get lost and the remaining six photons will be registered as a six-fold detector click in the output modes. Thus, $\dicke{6}{3}$ is mixed with highly colored noise, which exhibits different types of entanglement itself depending on the loss type. Insight into the coherence between the observed coincidences can be obtained from measurements in the $x$ [\fref{fig:counts}(b)] and $y$ [\fref{fig:counts}(c)] bases. The state \dicke{6}{3} transforms in these bases to $\sqrt{5/8}\ket{\mathrm{GHZ}_6^{\mp}}+\sqrt{3/16}(\dicke{6}{4}\mp\dicke{6}{2})$ with $\ket{\mathrm{GHZ}_N^{\mp}}=1/\sqrt{2}(\ket{0}^{\otimes N}\mp\ket{1}^{\otimes N})$ and $0=\{+,L\}$, $1=\{-,R\}$. In the experiment we observe the GHZ contribution as pronounced coincidence counts for the left- and rightmost projector. The residual counts from other terms [insets of \fref{fig:counts}(b) and (c)] make the decisive difference to a GHZ state as they are in a superposition with the GHZ terms. Apart from this, noise on top of all counts is also apparent. Most importantly, while the GHZ state shows its two pronounced terms only in a single basis, we observe these features now for two bases, which is directly related to the symmetry of \dicke{6}{3}.

\begin{figure}
\includegraphics{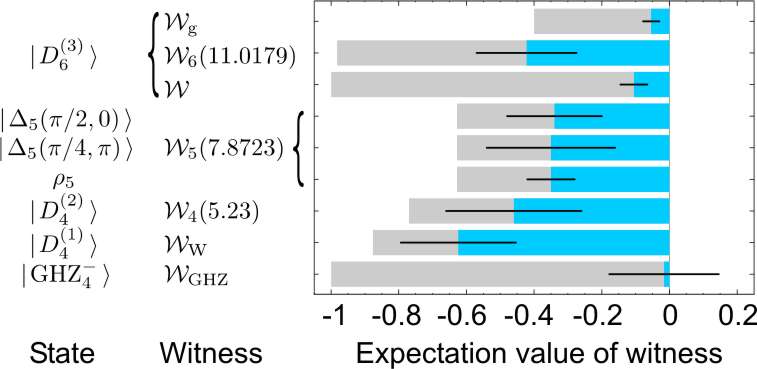}
\caption{\label{fig:witness}(color online) Experimental results (blue) and theoretical predictions (gray) are shown for the various entanglement witnesses for different states (see text). Negative values prove genuine $N$-partite entanglement.}
\end{figure}

A quantitative measure, indicating how well we prepared $\dicke{6}{3}$ experimentally, is given by the fidelity $F_{\Dicke{6}{3}}(\rho)=\mathrm{Tr}(\ket{\Dicke{6}{3}}\bra{\Dicke{6}{3}}\rho)$. Its determination would require 183 correlation measurements in the standard Pauli bases. However, employing the permutational symmetry of the state \dicke{6}{3} leads to a reduction to only 21 measurement settings \cite{Tot09tbd,Wie09NoteD63Fid}. We have determined $F_{\Dicke{6}{3}}=0.654\pm0.024$ with a measurement time of $31.5$\,h. This allows to apply the generic entanglement witness \cite{Tot07} $\langle \mathcal{W}_g\rangle=0.6-F_{\Dicke{6}{3}}=-0.054\pm0.024$ and thus proves genuine six-qubit entanglement of the experimentally observed state with a significance of two standard deviations (\fref{fig:witness}).

Proving entanglement based on witness operators can be much simpler in terms of the number of measurement settings, as due to the symmetry of $\dicke{6}{3}$ already the two measurements $x$ and $y$ are sufficient \cite{Kie07,Tot07,Tot07Spin}. The generic form of such a witness is given by $\mathcal{W}_N(\alpha)=\alpha\cdot{\openone}^{\otimes N}-({J_{N,x}}^2+{J_{N,y}}^2)$, where $\alpha$ is obtained by numerical optimization over all bi-separable states. For the state $\dicke{6}{3}$ $\mathcal{W}_6(11.0179)$ \cite{Tot09tbd} has a minimal value of $-0.9821$. In our experiment we have obtained with the data shown in \fref{fig:counts}(b) and (c) $\langle \mathcal{W}_6(11.0179)\rangle=-0.422\pm0.148$, i.e., after a measurement time of only $17.1$\,h a higher significance for proving six-qubit entanglement compared to the generic witness (\fref{fig:witness}). A different witness, allowing additionally to estimate the fidelity and requiring three measurement settings only, can be obtained by considering higher moments of the $J_{6,i}$ operators and is given as $\mathcal{W}=1.5\cdot\openone^{\otimes 6}-\sum_{i=x,y,z}\sum_{j=1}^3c_{ij}{J_{6,i}}^{2j}$ \cite{Tot09tbd}, with $c_{ij}=$ $(-1/45,1/36,-1/180;$ $-1/45,1/36,-1/180;$ $1007/360,-31/36,23/360)$. Experimentally, using the three measurements of \fref{fig:counts} we obtain $\langle \mathcal{W}\rangle=-0.105\pm0.040$ yielding also a quite accurate bound on the fidelity \cite{Tot09tbd} of $F_{\Dicke{6}{3}}\geq 0.6-\langle \mathcal{W}\rangle/2.5=0.642\pm0.016$ (\fref{fig:witness}). 

Another known method to reveal entanglement and additionally the non-classical nature of a quantum state are Bell inequalities. Introduced with the aim to exclude a local-realistic description of measurement results \cite{Bel64,Mer90PRL}, they recently became important tools in quantum information processing, e.g., for security analysis \cite{Sho00} or for state discrimination \cite{Sch08DS,Sch08DSX}. A Bell operator well suited for the latter task is given by, $\hat{\mathcal{B}}_{\Dicke{6}{3}}=4/5(\sigma_x\otimes M_5+\sigma_y\otimes M_5')$, where $M_5$ and $M_5'$ are five qubit Mermin operators \cite{Mer90PRL,Sch08DSX,Wie09NoteMermin5}.  The associated Bell inequality, $|\langle\hat{\mathcal{B}}_{\Dicke{6}{3}}\rangle_{\mathrm{avg}}|\leq0.4$, is maximally violated by the six-photon Dicke state ($\langle\hat{\mathcal{B}}_{\Dicke{6}{3}}\rangle_{\mathrm{\Dicke{6}{3}}}=1$) and much less, e.g., by any six-qubit GHZ state ($\langle\hat{\mathcal{B}}_{\Dicke{6}{3}}\rangle_{\mathrm{GHZ,max}}=0.85$). This again is a direct consequence of the particular symmetry of $\dicke{6}{3}$. While an inequality based on any of the two Mermin terms is maximally violated by a GHZ state, the violation of their sum is only maximal for $\dicke{6}{3}$ due to its symmetry and equal form in the $x$ and $y$ bases. The experimental value of $\langle\hat{\mathcal{B}}_{\Dicke{6}{3}}\rangle_{\mathrm{expt}}=0.43\pm0.02$ shows that there is no local realistic model describing this state, yet, due to the higher order SPDC noise, it is not sufficient to discriminate against GHZ states.

\begin{figure}
\includegraphics{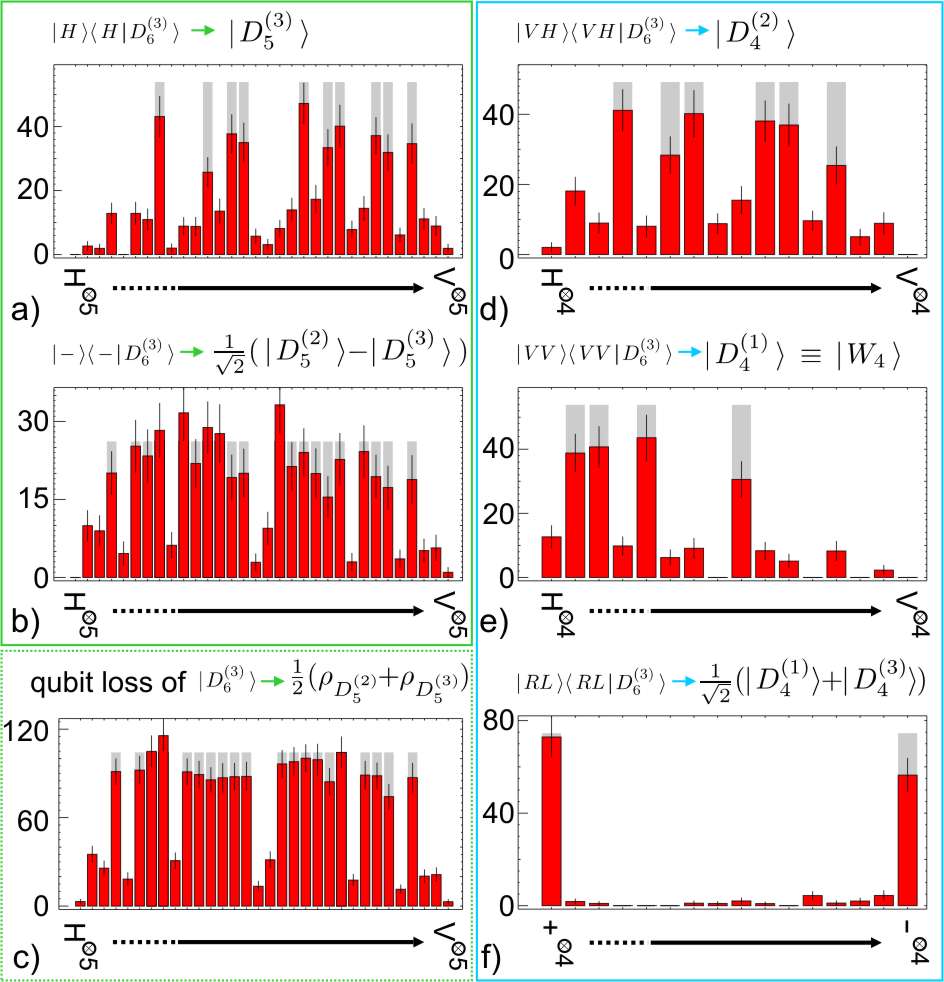}
\caption{\label{fig:projections}(color online) Experimentally measured coincidence counts in the $z$ basis [(a)-(e)] and $x$ basis [(f)] for projections of $\dicke{6}{3}$ to obtain (a)-(b) five, and (d)-(f) four qubit entangled states. (c) shows $\rho_5$ obtained after a loss of a qubit from $\dicke{6}{3}$. Each measurement took 279\,min. Theoretical predictions are shown as gray bars normalized to the total number of coincidences.}
\end{figure}

The characteristic symmetry and entanglement of \dicke{6}{3} enables one to observe a wealth of five- and four-qubit entangled states that can be obtained by projective measurements or qubit loss \cite{Wie09}. When we project one of the qubits onto $\cos{\theta}\ket{V}+\sin{\theta}e^{-i\phi}\ket{H}$, we first obtain superpositions of five qubit Dicke states, $\ket{\Delta_5(\theta,\phi)}=\cos{\theta}\dicke{5}{2}+\sin{\theta}e^{i\phi}\dicke{5}{3}$ with $\theta,\phi$ real. These states belong to two different SLOCC classes, one for the values $\theta=0$ or $\theta=\pi/2$ and the other one for the remaining value range \cite{Wie09}. \fref{fig:projections}(a) and (b) show measurements in the $z$ basis for a representative state of the two classes, either obtained by projecting a qubit onto $\ket{H}$ [$\ket{\Delta_5(\pi/2,0)}=\dicke{5}{3}$] or onto $\ket{-}$ [$\ket{\Delta_5(\pi/4,\pi)}=1/\sqrt{2}(\dicke{5}{2}-\dicke{5}{3})$]. \fref{fig:witness} shows measured expectation values of optimized entanglement witnesses for detecting genuine $N$-qubit entanglement of these and the following states. When a qubit of \dicke{6}{3} is {\emph{lost}} one obtains $\rho_5=1/2(\rho_{\Dicke{5}{2}}+\rho_{\Dicke{5}{3}})$, i.e., an equal mixture of $\dicke{5}{2}$ and $\dicke{5}{3}$ [\fref{fig:projections}(c)]. Remarkably and in sharp contrast to the case of loosing a qubit from a $\mathrm{GHZ}_6$ state, this mixed state is also genuine five-qubit entangled (\fref{fig:witness}). This fact now clearly provides, after all, a criterion to definitely distinguish these two prominent states and demonstrates the entanglement persistency \cite{Bou06} of \dicke{6}{3}.

By means of a second projective measurement we obtain a variety of SLOCC-inequivalent four qubit states. In \fref{fig:projections} we exemplarily show coincidences for three of those states. The state $\dicke{4}{2}$ \cite{Kie07} [\fref{fig:projections}(d)] is obtained by projection of one qubit onto $\ket{V}$ and another one onto $\ket{H}$. By projecting two qubits onto the same polarization (here $\ket{V}$) for the first time the four photon W state \cite{Zou02W4,Tas08}, i.e., $\dicke{4}{1}$, could be observed in a linear optics experiment [\fref{fig:projections}(e)]. Both states are clearly genuine four-partite entangled \cite{Kie07,Wie09NoteWitW4} as depicted in \fref{fig:witness}. We have determined fidelities of $F_{\Dicke{4}{2}}=0.682\pm0.022$ and $F_{\Dicke{4}{1}}=0.619\pm0.043$ using optimized measurement settings \cite{Tot09tbd,Guh07}. Possible applications of $\dicke{4}{1}$ and $\dicke{4}{2}$ comprise for example quantum telecloning, teleportation and secret-sharing \cite{Se03b,Mur99,Hil99,Kie07}. Most remarkably, one can also obtain a four-qubit GHZ-state, which is suitable for, e.g., secret sharing \cite{Hil99}. As mentioned before, there is a strong GHZ component in the state \dicke{6}{3}. Considering the representation in the $y$ basis [\fref{fig:counts}(c)], a projection of one photon onto $\ket{R}$ and another one onto $\ket{L}$ filters out just this GHZ component, but the remaining terms coherently superimpose to a four-qubit GHZ state, $\ket{\mathrm{GHZ}_4^-}=1/\sqrt{2}(\dicke{4}{1}+\dicke{4}{3})=1/\sqrt{2}(\ket{+}^{\otimes 4}-\ket{-}^{\otimes 4})$. The fourfold coincidence counts shown in \fref{fig:projections}(f) reveal the characteristic GHZ structure. However for this state a two-setting witness measurement \cite{Tot05PRL} resulted in a value of $\langle \mathcal{W}_{\mathrm{GHZ}}\rangle=-0.016\pm0.162$, which is not sufficient to prove entanglement with the relevant significance and can be attributed to the low fidelity of $F_{\mathrm{GHZ}}=0.528\pm0.042$ and the asymmetric GHZ structure [\fref{fig:projections}(f)].

Altogether, we have experimentally demonstrated in this letter remarkable entanglement properties of the Dicke state \dicke{6}{3}. It exhibits a high symmetry with characteristic correlations in various bases. As shown, this makes it a perfect resource for observing a wealth of different SLOCC-inequivalent states of a lower qubit number. The novel setup presented here allows experiments with a sufficient count rate and lays the foundations for demonstrations of important applications of \dicke{6}{3}, e.g., for phase-covariant telecloning, multi-partite quantum communication or entanglement enhanced phase measurements.

\begin{acknowledgments}
We would like to thank Christian Schmid, Wies{\l}aw Laskowski, Akira Ozawa and Thomas Udem for helpful discussions. We acknowledge the support of this work by the DFG-Cluster of Excellence MAP, the EU Project QAP and the
DAAD/MNiSW exchange program. W.W.~acknowledges support by QCCC of the Elite Network of Bavaria and the Studienstiftung des dt. Volkes. G.T. thanks the support of the National Research Fund of Hungary OTKA (T049234).
\end{acknowledgments}


\end{document}